\documentclass[a4paper,11pt]{article}
\usepackage{pos}

\usepackage{amsmath,amsfonts}
\usepackage{physics}
\usepackage{braket}

\usepackage{color}
\usepackage{comment}
\usepackage[comparestr,comparenum]{arraysort}

\usepackage{wrapfig}

\newcommand{\fix}[1]{\textcolor{red}{[#1]}}

\renewcommand{\logo}{\relax}

\title{A lattice regularization of Weyl fermions in a gravitational background
}

\author*[a]{Shoto Aoki}
\author[a]{Hidenori Fukaya}
\author[a]{Naoto Kan}

\affiliation[a]{Department of Physics, Osaka University,\\Toyonaka, Osaka 560-0043, Japan}
\emailAdd{saoki@het.phys.sci.osaka-u.ac.jp}
\emailAdd{hfukaya@het.phys.sci.osaka-u.ac.jp}
\emailAdd{kan@het.phys.sci.osaka-u.ac.jp}

\abstract{We report on a lattice fermion formulation with a curved domain-wall
mass term to nonperturbatively describe fermions in a gravitational background. In our
previous work in 2022, we showed
under the time-reversal symmetry that the edge-localized massless
Dirac fermion appears
on one and two-dimensional spherical domain-walls and the spin connection is induced on the lattice in a consistent way with continuum theory.
In this work, we extend our study to the Shamir type curved domain-wall fermions
without the time-reversal symmetry.
We find in the free fermion case that a single Weyl fermion appears
on the edge,
and feels gravity through the induced spin connection.
With a topologically nontrivial $U(1)$ gauge potential, however,
we find an oppositely chiral zero mode at the center where the gauge field is singular.
\par
Preprint number: OU-HET-1214
}

\FullConference{The 40th International Symposium on Lattice Field Theory (Lattice 2023)\\
July 31st - August 4th, 2023\\
Fermi National Accelerator Laboratory\\}

\begin{document}
\maketitle

\section{Introduction}



It is a challenge to formulate a gravitational background in lattice
gauge theory.
If one wants to express the curved metric by the lattice itself,
the link and site position of the lattice must be changed as a
flexible variables.
In the previous publications,  they employed a triangular lattice
to describe gravity classically or dynamically \cite{Hamber2009Quantum,Regge1961general,brower2016quantum,AMBJORN2001347Dynamicallytriangulating}
by varying the lengths and/or angles of the unit triangles.
It is, however, a nontrivial problem to take a continuum limit in a
systematic way
since there are ambiguities in how to increase the number of triangles,
smoothen the position-dependent angles, and shorten the links.
The rotational symmetry is, for instance, not automatically recovered
as in the case of the standard square lattice.


In \cite{Aoki:2022cwg,Aoki:2022aezanomalyinflow}, two of the authors proposed an alternative approach
to describe a lattice fermion system in a nontrivial gravitational background.
Instead of a curved lattice, a massive Dirac fermion is formulated in
the standard Euclidean square lattice.
The curved space is embedded as a domain-wall
where the mass term flips its sign.
This formulation was motivated by a mathematical theorem by Nash that
every curved
manifold can be embedded into some higher-dimensional Euclidean space
\cite{Nash1956TheImbedding}\footnote{
See similar attempts in condensed matter  theory \cite{Lee2009Surface,Imura2012Spherical,Parente2011Spin, Haller:2022pfj}
and experiments \cite{Szameit2010GeometricPotential,Onoe2012ObservationofRiemannian}.
}.
Then the metric as well as the vielbein and related connections
are uniquely determined by the embedding.
If the edge-localized modes appear on the curved domain-wall,
in the same way as the standard flat domain-wall fermion \cite{KAPLAN1992342AMethod}
they should feel gravity through the spin connections induced
by the Einstein's equivalence principle.




On the $S^1$ and $S^2$ curved domain-walls, we found that
the edge-localized massless Dirac fermion appears as an eigenstate of
the normal component of the Dirac gamma matrix,
the eigenvalue spectrum of the Wilson Dirac operator
contains an effect of the induced gravity, and a monotonous
recovery of the rotational symmetry in the naive classical continuum limit.
With the $U(1)$ gauge link variables, we also observed that the anomaly
inflow between bulk and edge is consistent with the Atiyah-Patodi-Singer index
theorem in the continuum theory.
Moreover, we succeeded in a microscopic description of the Witten effect
in terms of the curved domain-wall fermions \cite{aoki2021chiral, Aoki:2023lqp, Kan:2023latticepro}.

In this work, we extend our work in two directions.
One is to employ the Shamir domain-wall fermion formulation,
which ignores the positive mass region of the
Kaplan's construction \cite{Kaplan:2023pvd,Kaplan:2023pxd}.
Another extension is to consider a Weyl fermion on the wall,
which was not tried in \cite{Aoki:2022cwg,Aoki:2022aezanomalyinflow} where only time-reversal symmetric case was considered with massless Dirac edge modes.
We numerically investigate these new issues on
a three-dimensional square lattice having a $S^2$ domain-wall.

Our setup looks as if a single Weyl fermion sits on a single surface.
In the large radius limit of the sphere, one may expect
a possible lattice regularization of chiral gauge theory combining
the edge-localized Weyl fermions if the net gauge anomaly cancels.
A similar single-surface formulation was recently proposed in Refs. \cite{Kaplan:2023pvd,Kaplan:2023pxd}.
We, however, present an example that the low-energy effective theory is
not chiral but vector-like when the $U(1)$ gauge links have
a nontrivial topology.
This phenomenon may be an obstacle to formulating a chiral gauge theory.

\section{Shamir-type $S^2$ domain-wall in continuum theory}

We consider a three-dimensional disk $\mathbb{D}^3$ with radius $r_0$.
We assign a negative mass $-m$ on the disk and
$+M\to \infty$ outside $r=r_0$.
Then the system is a Shamir-type domain-wall with
a curved $S^2$ boundary around the disk. The Dirac operator $D$ with a $U(1)$ gauge connection $A_i$ and the hermitian conjugate are given by
\begin{align}
  D= &\sigma^i \qty(\pdv{}{x^i}-iA_i)-m,~
  D^\dagger= -\sigma^i \qty(\pdv{}{x^i}-iA_i)-m.
\end{align}

In this work, we investigate the eigenvalues/eigenfunctions of $D^\dagger D$ and $D D^\dagger $ rather than $D$ and $D^\dagger $ itself which are not Hermitian.
But in the polar coordinates, we need special care about the boundary condition at $r=r_0$. For two spinor fields $\psi_+$ and $\chi_+$ in a domain of $D^\dagger D$, the inner product is given by
\begin{align}
  (\psi_+, D^\dagger D \chi_+)-(D \psi_+,  D \chi_+)=-\qty[\int_{S^2} d\Omega^2 r^2 \psi_+^\dagger  \sigma_r D \chi_+ (r)]_{r=0}^{r=r_0},
\end{align}
where $\sigma_r=\frac{\sigma^j x^j}{r} $ is a Pauli matrix in the normal direction to the boundary and the solid angle $d\Omega^2$ integral on the right-hand side must vanish. We, therefore, impose {the regularity at $r=0$} and the following boundary condition
\begin{align}\label{eq:BC for DdaggerD}
  \begin{aligned}
    \lim_{r\to r_0}(\sigma_r  -1 )\psi_+ &= 0,&\lim_{r\to r_0}(\sigma_r  + 1 )D\psi_+ &= 0 
  \end{aligned}
\end{align} 
which eliminates the boundary term. Note that $\psi_+(r_0)$ and $D\chi_+(r_0)$ are orthogonal to each other. Similarly, we also introduce a boundary condition 
\begin{align}\label{eq:BC for DDdagger}
  \begin{aligned}
    \lim_{r\to r_0}(\sigma_r  +1 )\psi_- &= 0,&\lim_{r\to r_0}(\sigma_r  - 1 )D\psi_- &= 0 
  \end{aligned}
\end{align} 
for a spinor $\psi_-$ in a domain of $D D^\dagger $. The subscripts $\pm$ mean the chirality at $r=r_0$.

When $A_i=0$, the Dirac operator is written as
\begin{align}
  D= \sigma_r \qty(\pdv{}{r} +\frac{1}{r} -\frac{D^{S^2}}{r} ) -m,\quad  D^{S^2}=\sigma^a L_a  +1,
\end{align}
where $L_a=-i\epsilon_{abc} x_b \partial_c$ are angular momentum operators. $D^{S^2}$ and $ \sigma_r= \sigma^j x^j/r$ anti-commute with each other. By a local (Euclidean version of) Lorentz as well as an additional $U(1)$ rotation,
we obtain
\begin{align}\label{eq:transformation of DS2}
  sD^{S^2}s^{-1}=i\qty(\sigma_1 \pdv{}{\theta} +\frac{\sigma_2}{\sin \theta } \qty(  \pdv{}{\phi}-i A^{\text{Spin}^c}_\phi  ) ),\quad  
 s  \frac{\sigma^j x^j}{r}s^{-1}= \sigma^3,
\end{align}
where $s=\exp(i\theta \sigma_2/2 ) \exp(i\phi (\sigma_3+1)/2 ) $ and we can see a gravitational effect as a nontrivial spin$^c$ connection 
\begin{align} \label{eq:spin connection on S2}
  A^{\text{Spin}^c}_\phi= -\frac{1}{2} +\frac{\cos\theta}{2 } \sigma_3.  
\end{align}
Then, $D^{S^2}$ turns into the Dirac operator on $S^2$ and $\sigma_r$ becomes the chiral operator.

We solve eigenvalue problems of $D^\dagger D\psi_+ = E^2 \psi_+ $ and $D D^\dagger \psi_- = E^2 \psi_- $.
The total angular momentum operators $ J_a= L_a+ \frac{\sigma_a}{2}$ satisfy
\begin{align}\label{eq:commutation relation}
  [J_a, D]=0,~
  [J_a,D^\dagger]=0,~
  [J_a ,D^{S^2}]=0, ~
  [J_a, \sigma_r]=0,~
  [J_a,J_b]=i\epsilon^{abc} J_c, 
\end{align}
so $J_a$ commutes with $D$ and $D^\dagger$. Let $j$ and $j_3$ be an azimuthal quantum number and magnetic quantum number for $J_a$, then $j$ and $j_3$ take a value in
\begin{align}
  j= \frac{1}{2},~\frac{3}{2},\cdots,~
  j_3= -j,~-j+1,~\cdots,j.
\end{align}
We find a simultaneous eigenstate of $J_3 ,~J^2$ and $D^{S^2}$ as
\begin{align}
  J^2 \chi_{j,j_3,\pm}&=j(j+1) \chi_{j,j_3,\pm} ,~
  J_3 \chi_{j,j_3,\pm}=j_3 \chi_{j,j_3,\pm}\\
  D^{S^2} \chi_{j,j_3,\pm}&= \pm \nu \chi_{j,j_3,\pm},~
  \sigma_r \chi_{j,j_3,\pm}= \chi_{j,j_3,\mp},
\end{align}
where $\nu=j+1/2$ is an eigenvalue of the Dirac operator on $S^2$.

We obtain the edge localized mode written as the linear combination of $\chi_{j,j_3, \pm}$:
\begin{align}\label{eq:DW localized mode}
  \psi_{j,j_3,\pm}^\text{DW}= \frac{1}{\sqrt{r}} \qty(  \sqrt{m-E} I_{\nu-1/2}(\sqrt{m^2-E^2}r) \chi_{j,j_3,+}\pm  \sqrt{m+E} I_{\nu+1/2}(\sqrt{m^2-E^2}r)\chi_{j,j_3,-} ).
\end{align} 
for $\abs{E}<m$. 
Note that $(2j + 1)$ states share the same value of $E$. From the first boundary condition, $E$ satisfy
\begin{align} \label{eq:E continuum}
  \frac{ I_{\nu-1/2}(\sqrt{m^2-E^2}r_0) }{ I_{\nu+1/2}(\sqrt{m^2-E^2}r_0)}= \frac{ \sqrt{m+E} }{ \sqrt{m-E}}
\end{align}
and the second condition is automatically held. The boundary condition for $r=r_0$ is obviously satisfied from this expression. This state is localized at the boundary $r=r_0$ exponentially. In the large $m$ limit, $E$ converges to
 \begin{align}
  E \to \frac{\nu}{r_0} 
\end{align}
and $\psi_{j,j_3,\pm}^\text{DW}$ becomes a Weyl fermion with background gravity on $S^2$. It seems that $D^\dagger D$ and $DD^\dagger $ define a Weyl fermion system on $S^2$ whose chirality is positive and negative, respectively.


In the following, we consider a spherically symmetric $U(1)$ gauge field.
Such a background can be given by
\begin{align} \label{eq:gauge conn n-monopole}
  A_\phi^{U(1)}=n\frac{1-\cos \theta}{2}, 
\end{align}
which is equivalent to putting a magnetic monopole with a magnetic charge $n$
at the center\footnote{
To be precise, it is not a magnetic monopole in three-dimensional space but
it is a solitonic object in $(2+1)$-dimensional Euclidean spacetime.
}. Then the $D^{S^2}$ and angular momentum operators $L_1,~L_2$ and $L_3$ are modified by
\begin{align}
  D^{S^2}=\sigma^a \qty(L_a +\frac{n}{2} \frac{x_a}{r} )+1,~L_a=-i\epsilon_{abc} x_b (\partial_c -iA_c)-\frac{n}{2} \frac{x_a}{r}.
\end{align}
By the same transformation in the equation \eqref{eq:transformation of DS2},~$D^{S^2}$ becomes the Dirac operator on $S^2$ with the gauge field:
\begin{align}\label{eq:Dirac op on S2 mono}
  D^{S^2} &\to sD^{S^2}s^{-1}=i\qty(\sigma_1 \pdv{}{\theta} +\frac{\sigma_2}{\sin \theta } \qty(  \pdv{}{\phi} -iA^{\text{Spin}^c}_\phi  -i A^{U(1)}_\phi ) ).
\end{align}
The commutation relations are the same as the equation \eqref{eq:commutation relation} but the azimuthal quantum number $j$ and the eigenvalue $\nu$ of $D^{S^2}$ change to 
\begin{align}
  j= \frac{\abs{n}-1}{2},~ \frac{\abs{n}+1}{2},\cdots,~\nu=\sqrt{\qty(j+\frac{1}{2})^2 -\frac{n^2}{4} }.
\end{align} 

When $j\geq \frac{\abs{n}+1}{2}$, $\nu $ is not zero and we get the same $\chi_{j,j_3,\pm}$ as \eqref{eq:DW localized mode}. Thus the localized mode is also expressed by $\psi^{DW}_{j,j_3}$.

When $j= \frac{\abs{n}-1}{2}$, $\nu$ is equal to zero and the situation is different from the above case. The corresponding state is also an eigenstate of $\sigma_r$ so the simultaneous eigenstate is obtained as
\begin{align}
  J^2 \chi_{j,j_3,0}=j(j+1) \chi_{j,j_3,0} ,~&
  J_3 \chi_{j,j_3,0}=j_3 \chi_{j,j_3,0}\\
  D^{S^2} \chi_{j,j_3,0}= 0,~
  &\sigma_r \chi_{j,j_3,0}= \text{sign}(n)\chi_{j,j_3,0}.
\end{align}
The eigenstate of $D^\dagger D$ and $DD^\dagger$ are described as 
\begin{align}
  \psi_{j,j_3,\pm} =   \frac{1 \pm \text{sign}(n) }{2r} \sinh \kappa r  \chi_{j,j_3,0}
\end{align}
and degenerates $(2j+1)=\abs{n}$-fold. By the boundary conditions, $\kappa$ must satisfy
\begin{align}
  \tanh (\kappa r_0)= \frac{\kappa}{ m}
\end{align}
so the eigenvalue slightly shifts from $E=0$. If $m$ is larger than $1/r_0$, there is a solution. In the large mass limit, the $\kappa $ converges to $m$ and 
\begin{align}
  E=\frac{\sqrt{m^2-\kappa^2}}{r_0} \to 0.
\end{align}
{This state appears when $\pm \text{sign}(n)$ is positive and is localized at the wall.}

\section{Shamir-type $S^2$ domain-wall on a lattice}

\begin{wrapfigure}[14]{r}[0pt]{0.4\textwidth}
  \centering
  \includegraphics[bb= 0 0 230 229, width =0.4\textwidth]{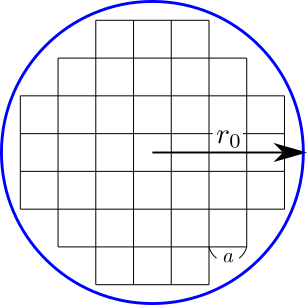}
  \caption{A two-dimensional slice of the three-dimensional square lattice having the $S^2$ domain-wall at $r=r_0$.}
  \label{fig:ShamirTypeDW}
 \end{wrapfigure}

We discretize the three-dimensional disk $\mathbb{D}^3$ with the radius $r_0$ and make a square lattice space with the lattice spacing $a$ (see Figure \ref{fig:ShamirTypeDW}). The lattice point is expressed by $x=(x^1,x^2,x^3)$ and $\sum_i (x^i)^2<r_0^2$. $x^i$ takes a value in $x^i=\cdots, -a/2,+ a/2,\cdots$. 
Note that the origin is put at x=0, which is not any site position of the lattice.
On this lattice we consider the Wilson-Dirac operator 
\begin{align}\label{eq:Wilson Dirac op on S2}
  D =\frac{1}{a}\qty(\sum_{i=1}^3\qty[\sigma_i\frac{\nabla_i-\nabla^\dagger_i}{2} +\frac{1}{2}\nabla_i \nabla^\dagger_i ]- am ), 
\end{align}
\noindent where $\nabla_i $ is the covariant difference operator in the $i$-direction and $\nabla_i^\dagger $ is a hermitian conjugate of $\nabla_i$:
\begin{align}
  (\nabla_i \psi)_x &=\exp( -i \int_{x+\hat{i}}^x A  ) \psi_{x+\hat{i}}-\psi_{x}, \\
  (\nabla_i^\dagger \psi)_x &=\exp( -i \int_{x-\hat{i}}^x A  ) \psi_{x-\hat{i}}-\psi_{x}.
\end{align}
The sigma matrix in the $r$-direction $\sigma_r=\frac{\sigma^j x^j}{r}$ is well-defined on the square lattice space since $x=0$ is not a lattice point.

\begin{figure}[b]
\begin{minipage}[b]{0.45\linewidth}
\centering
\includegraphics[bb=0 0 521 316 , width=\textwidth ]{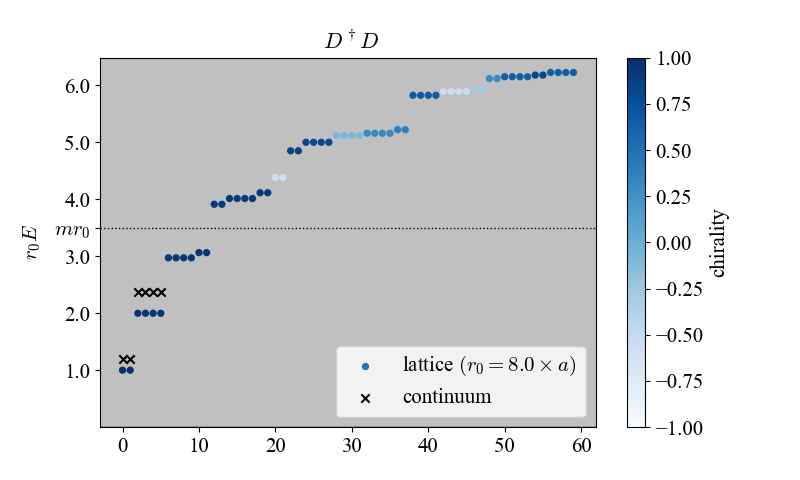}
\end{minipage}
\hfill
\begin{minipage}[b]{0.45\linewidth}
  \centering
  \includegraphics[bb=0 0 521 316 , width=\textwidth ]{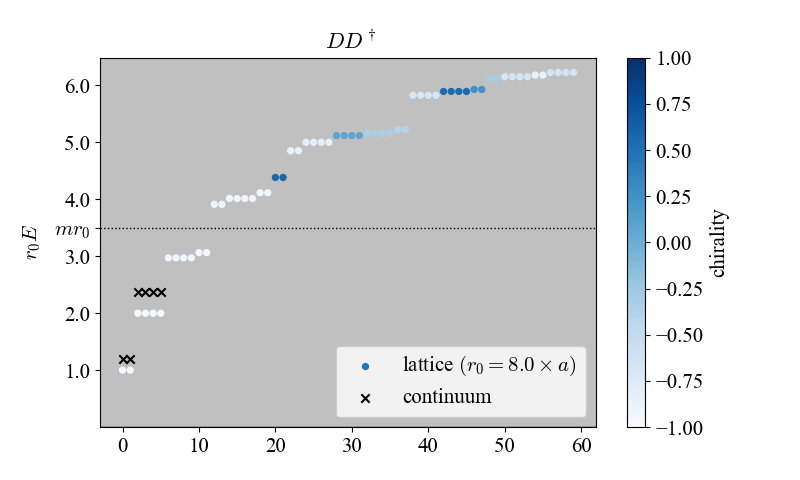}
  \end{minipage}
\caption{Left panel: The eigenvalue spectrum of $D^\dagger D$ when $mr_0=3.5,~r_0=8a$ and $n=0$. Right: That of $DD^\dagger$. The color gradation represents their chirality, which is the expectation value of $\sigma_r$. The cross symbols denote a continuum prediction \eqref{eq:E continuum}.}
\label{fig:spectrum n=0}
\end{figure}

\begin{figure}
  \begin{minipage}[b]{0.45\linewidth}
  \centering
  \includegraphics[bb=0 0 521 316 , width=\textwidth ]{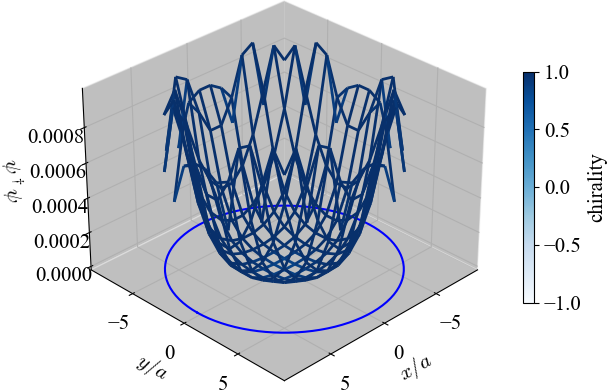}
  \end{minipage}
  \hfill
  \begin{minipage}[b]{0.45\linewidth}
    \centering
    \includegraphics[bb=0 0 521 316 , width=\textwidth ]{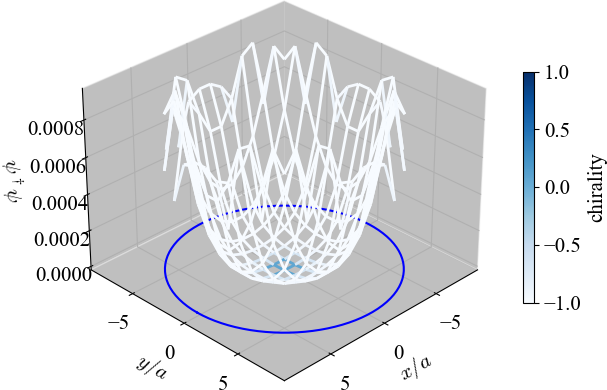}
    \end{minipage}
  \caption{Left panel: The amplitude of an eigenmode of $D^\dagger D$ with $E=1/r_0$ on $z=a/2$ plain when $mr_0=3.5,~r_0=8a$ and $n=0$. Right: That of $DD^\dagger$. The color represents the chirality on each lattice point. The blue circle on the bottom is the location of the $S^2$ domain-wall.
  }
  \label{fig:amplitude n=0}
\end{figure}

We analyze the eigenvalue problem of $D^\dagger D$ and $DD^\dagger$ at $mr_0=3.5 $ and $r_0= 8a$, numerically. In the absence of the monopole $(n=0)$, we plot the eigenvalue of $D^\dagger D$ and $D D^\dagger$ as filled circle symbols in Figure \ref{fig:spectrum n=0}. The crosses are a continuum result predicted by \eqref{eq:E continuum}. The color gradation represents the chirality, which is the expectation value of $\sigma_r$. The darker it is, the closer it is to $+1$. 
We plot the amplitude of an eigenfunction with $E=1/r_0$ in Figure \ref{fig:amplitude n=0}. The color gradation shows the chirality on each lattice point.
They are localized at the boundary.
The gap in the eigenvalue spectrum agrees well with the continuum prediction
having the induced spin connection in Eq. \eqref{eq:spin connection on S2}.
This indicates that the edge-modes of the lattice curved domain-wall fermion
feels gravity induced on the curved surface even though it is put on
the square lattice.


Our fermion system has a single domain-wall.
The edge modes have a definite single chirality.
Therefore, it looks as if this lattice curved domain-wall fermion
offers a non-perturbative regularization of a  Weyl fermion.
However, with a $U(1)$ gauge field having a non-trivial topological charge, we find a numerical situation contrary to this expectation.

\begin{figure}
  \begin{minipage}[b]{0.45\linewidth}
  \centering
  \includegraphics[bb=0 0 521 316 , width=\textwidth ]{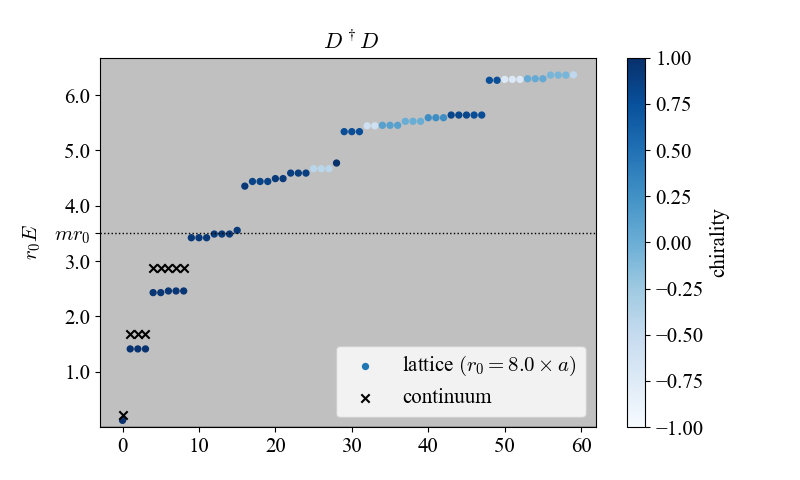}
  \end{minipage}
  \hfill
  \begin{minipage}[b]{0.45\linewidth}
    \centering
    \includegraphics[bb=0 0 521 316 , width=\textwidth ]{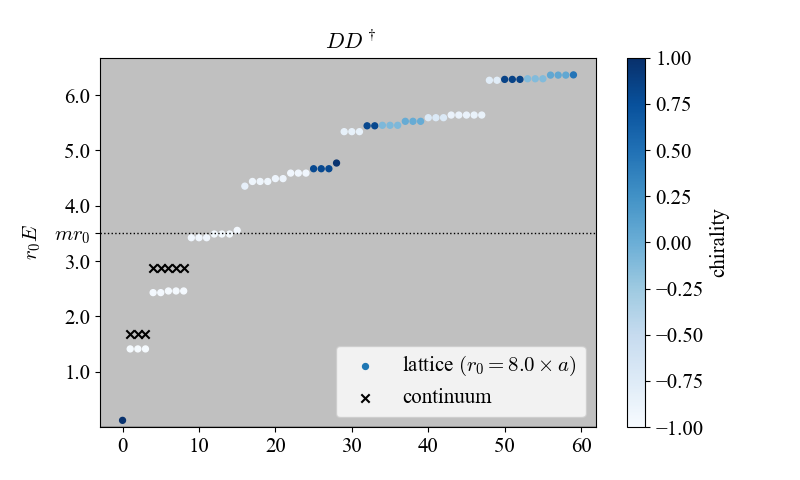}
    \end{minipage}
    \caption{Left panel: The eigenvalue spectrum of $D^\dagger D$ when $mr_0=3.5,~r_0=8a$ and $n=1$. Right: That of $DD^\dagger$. The color gradation represents their chirality, which is the expectation value of $\sigma_r$. The cross symbols denote a continuum prediction \eqref{eq:E continuum}.}
  \label{fig:spectrum n=1}
\end{figure}

\begin{figure}
\begin{minipage}[b]{0.45\linewidth}
\centering
\includegraphics[bb=0 0 521 316 , width=\textwidth ]{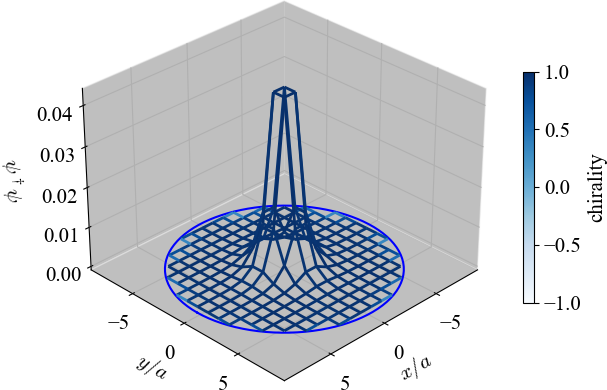}
\end{minipage}
\hfill
\begin{minipage}[b]{0.45\linewidth}
  \centering
  \includegraphics[bb=0 0 521 316 , width=\textwidth ]{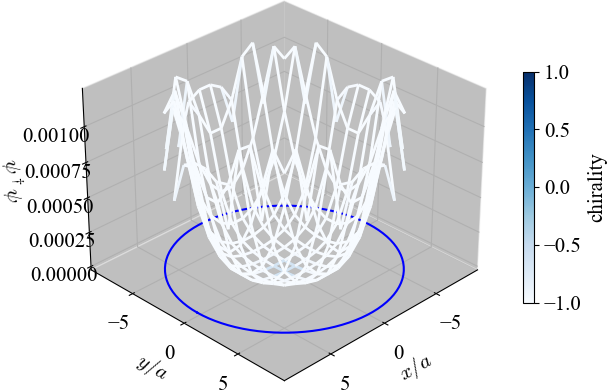}
\end{minipage}
\caption{Left panel: The amplitude of the zero-mode of $DD^\dagger $ on $z=a/2$ plain when $mr_0=3.5,~r_0=8a$ and $n=1$. Right: That of the first excitation mode of $DD^\dagger$. The color represents the chirality on each lattice point. The blue circle on the bottom is the location of the $S^2$ domain-wall.}
\label{fig:amplitude new zeromode}
\end{figure}

\begin{figure}
  \begin{minipage}[b]{0.45\linewidth}
  \centering
  \includegraphics[bb=0 0 521 316 , width=\textwidth ]{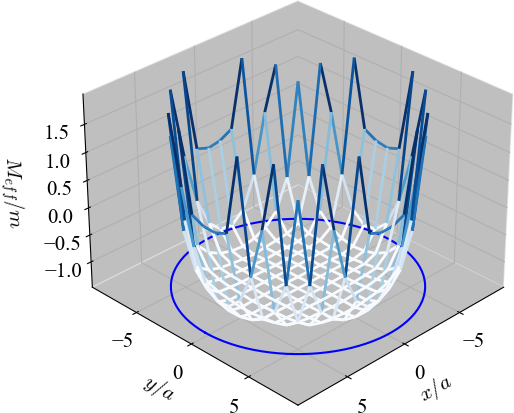}
  \end{minipage}
  \hfill
  \begin{minipage}[b]{0.45\linewidth}
    \centering
    \includegraphics[bb=0 0 521 316 , width=\textwidth ]{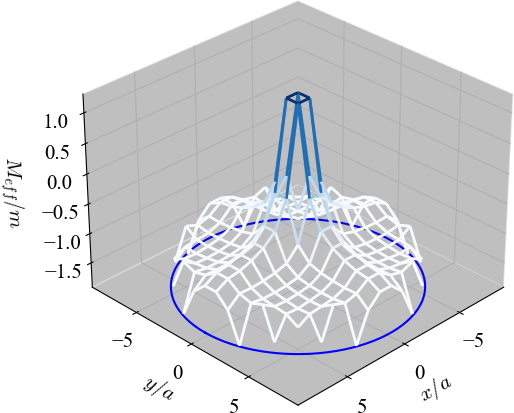}
    \end{minipage}
  \caption{Left panel: The amplitude of the effective mass for the lowest eigenvalue mode of $DD^\dagger$ on $z=a/2$ plain when $mr_0=3.5,~r_0=8a$ and $n=0$. Right: That of $n=1$. The color represents the magnitude on each lattice point. The blue circle on the bottom is the location of the $S^2$ domain-wall.
  }
  \label{fig:effective mass}
\end{figure}

We numerically solve the eigenvalue problem when $n = 1$ and plot the eigenvalue
spectrum in Figure \ref{fig:spectrum n=1}. 
{The spectrum of $D^\dagger D$ is consistent with the continuum
prediction but that for $DD^\dagger$ is not: another zero mode with the opposite
chirality appears.}
As shown in Fig.~\ref{fig:amplitude new zeromode}, this oppositely chiral zero mode is located at the center,
where the gauge field is singular.
In fact, the singularity of the gauge field gives a strongly additive
mass renormalization through the Wilson term to make the effective
mass positive (See Figure \ref{fig:effective mass}).
This dynamical creation of the domain-wall is the origin of the
oppositely chiral zero mode.


From a cobordism property discussed in our previous work \cite{Aoki:2023lqp},
the singularity of the gauge field and accompanying dynamical creation
of the domain-wall
seems to be unavoidable when the gauge field is topologically nontrivial.
The paring of the $\pm$ chirality in the zero modes indicate that the
low-energy effective
theory is not chiral but vectorlike.
It will be an interesting study to examine if we can eliminate the
unwanted chiral zero modes for instance, 
by a symmetric mass generation \cite{Kikukawa:2017ngf}
when the target theory on the single domain-wall is anomaly free \cite{Kaplan:2023pvd,Kaplan:2023pxd}.


\section{Summary}

We have considered a massive Dirac fermion on a three-dimensional square lattice
with a spherically curved domain-wall. By cutting off the hopping to
the outside of the surface, the curved domain-wall fermion is of the
Shamir type.
In the free theory case, we have confirmed appearance of a single Weyl
fermion on the domain-wall.
Moreover, our numerical results for the Dirac eigenvalue spectrum
shows a gravitational effect consistent with  the expected spin
connection predicted in continuum theory.
With a topologically nontrivial gauge field background, however,
we have observed emergence of the oppositely chiral zero mode
at the center of our system where the gauge field is singular.
We have confirmed that this new zero-mode sits on a dynamically
created small domain-wall,
which makes the low energy theory not chiral but vectorlike.


\acknowledgments

We thank S. Aoki, M. Furuta, S. Iso, D.B. Kaplan, Y. Kikukawa, M. Koshino, Y. Matsuki, S. Matsuo, T. Onogi, S. Yamaguchi, M. Yamashita and R. Yokokura for useful discussions. 
The work of SA was supported by JSPS KAKENHI Grant Number JP23KJ1459. The work of HF and NK was supported by JSPS KAKENHI Grant Number JP22H01219.

\typeout{} 
\bibliographystyle{JHEP}
\bibliography{ref.bib}

\end{document}